\documentclass{appolb}
\usepackage{epsfig}
\usepackage{amssymb}
\usepackage{amsmath}
\usepackage{subfigure}
\usepackage{graphics}
\usepackage{graphicx}
\newcommand{\dd}{\mbox{\rm d}}

\newcommand{\pol}[1]{\mathaccent"017E{#1}}

\begin{document}
\date{\today}
\pagestyle{plain}
\newcount\eLiNe\eLiNe=\inputlineno\advance\eLiNe by -1
\title{AN INTRODUCTION TO MESIC NUCLEI%
\thanks{Presented at the Symposium on Fundamental and Applied
Subatomic Physics, Krak\'ow, June 7---12, 2015}%
}
\author{Colin Wilkin
\address{Physics and Astronomy Dept., UCL, Gower Street, London, WC1E
6BT, UK }} \maketitle

\begin{abstract}
There is much speculation and a modest amount of evidence that certain mesons
might form quasi-bound states with nuclei to produce really exotic states of
matter. For this to be a practical possibility, the interaction between the
meson and nucleons at  low energies must be strong and attractive and the
production rates ``healthy''. The conditions for this are surveyed for the
$\bar{K}$, $\eta$, $\omega$, $\eta^{\prime}$, and $\phi$ mesons. How this
might lead to quasi-bound states is then discussed in a few typical cases.

Though some interesting effects have been seen in above-threshold data, the
search for experimental signals for these exotic states with different mesons
in bound state regions has generally been rather frustrating, with positive
claims only being made for the $\eta$ and the $K^-$.
\end{abstract}

\section{Introduction}
\label{Introduction}

Though the field had been around for more than 20 years, in 2010 the
APS finally recognised that there was a subject called ``mesic nuclei''.
\begin{center}
\fbox{PACS 2010: 21.85.+d Mesic nuclei}
\end{center}

The subject is of interest for a variety of reasons:
\begin{itemize}
\item It allows one to investigate the interaction of unstable particles with
nucleons and nuclei.
\item If such states existed, they would represent exotic nuclear matter with
several hundred MeV of excitation energy.
\item If one can produce a $_{\eta}^3$He through $dp \to{} _{\eta}^3$He,
    this will contribute a small amount to $dp \to{} _{\eta}^3$He${}\to
    dp$, \textit{i.e.}, deuteron-proton elastic scattering.
\end{itemize}

\section{Near-threshold production of mesons in elementary reactions}

There is ample evidence from many sources that the $s$-wave $\pi^0 p$
interaction is very weak and it stays weak until the $p$-wave $\Delta(1232)$
is approached. Nobody has therefore serious hopes of exotic nuclei involving
pions and so we turn immediately to heavier mesons.

\subsection{$\eta$ and $\eta^{\prime}$ production}

If one only includes the $S$-wave and its final state interaction (FSI) in
the $pp$ system in the simplest possible approximation~\cite{Faldt_96}, one
expects the $pp\to pp\eta$ total cross section to vary as
\begin{equation}
\label{FW1}
\sigma_T(pp\to pp\eta)=C \left.\left(\frac{Q}{\varepsilon}\right)^2\right/
   \left(1+\sqrt{1+Q/\varepsilon}\right)^{2},
\end{equation}
where the excess energy $Q=W-(2m_p+m_{\eta})c^2$, with $W$ being the total
c.m.\ energy. The constant $C$ depends upon the reaction mechanism and can be
adjusted to fit the data.

\begin{figure}[htb]
\begin{center}
\includegraphics[width=0.55\textwidth]{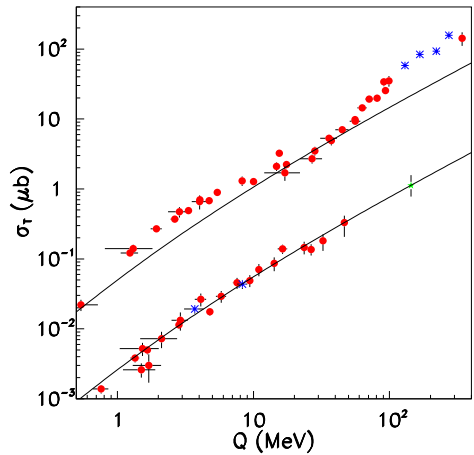}
\caption{\label{fig:eta_etaprime} Total cross sections for $pp\to pp\eta$
(upper points) and $pp\to pp\eta^{\prime}$ (lower points). The $\eta$ data
are taken from
Refs.~\cite{Bergdolt_93,Hibou_98,Chiavassa_94a,Calen_96,Smyrski_00,Moskal_04,Moskal_10}
(closed red circles), and \cite{Marco_01} (blue crosses) and the
$\eta^{\prime}$ data from Ref.~\cite{Bergdolt_93,Hibou_98} (blue crosses),
\cite{Balestra_00} (green star), and \cite{Moskal_98,Moskal_00,Czerwinski_14}
(closed red circles). The solid curves are  arbitrarily scaled $pp$ FSI
predictions of Eq.~(\ref{FW1}). }
\end{center}
\end{figure}

Since the Coulomb repulsion has here been
neglected, there is some ambiguity in the value to take for the pole position
$\varepsilon$. The best fit to the analogous $\eta^{\prime}$ production
data~\cite{Bergdolt_93,Hibou_98,Balestra_00,Moskal_98,Moskal_00,Czerwinski_14}
was achieved with $\varepsilon =
0.75^{+0.20}_{-0.15}$~MeV~\cite{Czerwinski_14}, which is quite consistent
with the original theoretical assumptions~\cite{Faldt_96}. The resulting
curves for $\eta$ and $\eta^{\prime}$ production are compared to experimental
data in Fig.~\ref{fig:eta_etaprime}.

In the $\eta$ case there are large deviation from the curve at low $Q$ that
may be ascribed to a strong $\eta$-nucleon FSI. The deviations at large $Q$
are likely to originate from $P$ or higher waves in the final $pp$ system.
The situation is very different for $\eta^{\prime}$ production, where there
is no sign of any FSI in the near-threshold data. The COSY-11 collaboration
has put limits on the $\eta^{\prime}p$ scattering length but one can see
immediately from the figure that the FSI is very much stronger for $\eta$
production. One cannot make statements regarding the influence of $P$-waves
for the $\eta^{\prime}$ at large $Q$ due to the lack of data there. The other
point that is worth noting is the factor of about twenty between $\eta$ and
$\eta^{\prime}$ production cross sections.

\subsection{$\omega$ production}

\begin{figure}[htb]
\begin{center}
\includegraphics[width=0.55\textwidth]{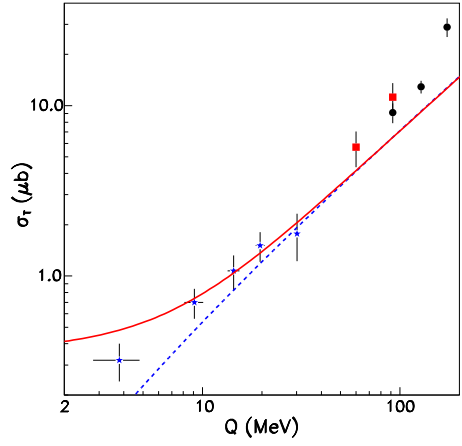}
\caption{\label{fig:omega1} Total cross sections for $pp\to pp\omega$ in
terms of the nominal value of $Q$, i.e., neglecting the $\omega$ width. The
data are taken from Refs.~\cite{Hibou_99} (blue crosses), \cite{Barsov_07}
(red squares), and \cite{Bary_10} (black circles). The (blue) dashed curve is
an arbitrarily scaled $pp$ FSI prediction of Eq.~(\ref{FW1}), whereas the
(red) solid one has been smeared over the $\omega$ width. }
\end{center}
\end{figure}

The situation is much less certain for $\omega$ production. The $pp\to
pp\omega$ total cross section is of the same order of magnitude as that for
$\eta$ production but, because of its natural width
$\Gamma_{\omega}\approx8.5$~MeV/$c^2$, the missing-mass peak is generally
less narrow. The comparison of the data in Fig.~\ref{fig:omega1} with the
predictions of Eq.~(\ref{FW1}) looks very similar to that of the $\eta$ in
Fig.~\ref{fig:eta_etaprime}. The strong deviations seen at high $Q$ are
likely to originate from contributions of $P$ and higher waves in the final
$pp$ system. At low $Q$ the points also lie above the (dashed) $pp$ FSI
curve. This is (probably) not due to any attraction between the $\omega$ and
a proton but rather it is an effect of the natural $\omega$ width. Even when
the nominal $Q$ is negative, one can still produce the low mass tail of the
$\omega$. The solid (red) curve tries to take this into
account~\cite{Hibou_99}.

\begin{figure}[htb]
\begin{center}
\includegraphics[width=0.55\textwidth]{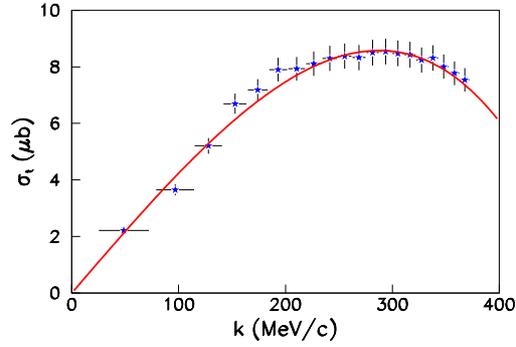}
\caption{\label{fig:Igor} The total cross section for $\gamma p \to p\omega$
as a function of the nominal (zero $\omega$ width) value $k$ of the final
c.m.\ momentum~\cite{Strakovsky_15}. The data are compared to a fit of the
form $\sigma_{t}= a_1k+a_3k^3+a_5k^5$.}
\end{center}
\end{figure}

Since it is hard to draw conclusions on the $\omega p$ FSI from the $pp\to
pp\omega$ data of Fig.~\ref{fig:omega1}, let us turn instead to the new
near-threshold data on $\gamma p \to
p\omega$~\cite{Strakovsky_15}\footnote{New data on this reaction have
appeared post-symposium from CBELSA~\cite{Wilson_15}.}. Smearing over the
$\omega$ decay width is less critical here because the unsmeared cross
section rises far more rapidly from threshold than it does for $pp\to
pp\omega$. The threshold value is $\sigma_t/q \approx 0.044~\mu$b/MeV/$c$. In
the vector-dominance model, the photon reactions are related to ones driven
by incident $\rho$, $\omega$, and $\phi$ vector mesons, from which one can
get an estimate of the $\omega p$ scattering length, $a_{\omega p} = (0.82\pm
0.03)$~fm. However, one doesn't really know how much of this is due to the
$\rho$-meson or the Born term.

Theoretical models give typically much smaller values of the scattering
length, e.g., $a_{\omega p} = (-0.026+i0.28)$~fm from the coupled-channel
analysis of $\omega$ production in $\gamma N$ and $\pi N$
interactions~\cite{Shklyar_05}, but there lots of other models on the market.

\subsection{$\phi$ production}

\begin{figure}[htb]
\begin{center}
\includegraphics[width=0.5\textwidth]{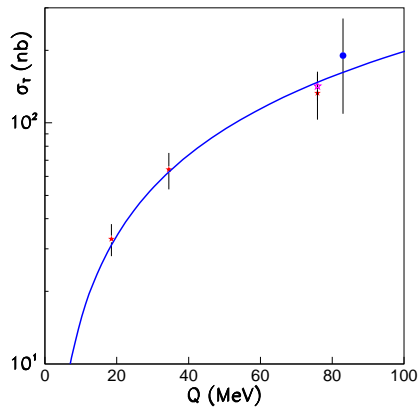}
\caption{\label{fig:phi} The total cross section for $pp\to pp\phi$ measured
at ANKE (red stars)~\cite{Hartmann_06} and DISTO (blue
circle)~\cite{Balestra_01}. The curve is the simple $pp$ FSI parameterisation
of Eq.~(\ref{FW1}). For this range of $Q$ the smearing over the $\phi$ width
is not important.}
\end{center}
\end{figure}

The available data on the $pp\to pp\phi$ total cross section are shown in
Fig.~\ref{fig:phi}, along with the curve corresponding to the simple $pp$ FSI
approach of Eq.~(\ref{FW1}). The good agreement with the curve may be
completely fortuitous because the data show that many partial waves must be
present at the higher excess energies. It should, however, be noted that the
cross section is about a factor of 30 lower than that for $\omega$
production. There is a lack of data at small $Q$ and so it is not clear from
these results whether the $s$-wave $\phi p$ interaction is attractive or not.

It may seem surprising that the energy dependence of $\phi$ production is
better measured in the $pn \to d\phi$ reaction~\cite{Maeda_06}. The message
is, however, similar in that the total cross section behaves like $\sqrt{Q}$,
i.e., like phase space with no sign of any $\phi$ attraction to the deuteron
at low energy. One should, nevertheless, bear in mind that the $\phi$ decay
distribution shows higher partial waves above about 40~MeV.

\subsection{$K^-$ production}

Due to strangeness conservation, a $K^-pp$ or $K^0pp$ system cannot be
produced in isolation in $pp$ collisions. The best that can be done is to
look at kaon pair production, $pp\to K^+K^-pp$, and compare the $K^+pp$ and
$K^-pp$ distributions. The $K^+$ is believed to be weakly interacting with
nucleons and the force may even be slightly repulsive!

\begin{figure}[htb]
\begin{center}
\includegraphics[width=0.52\textwidth]{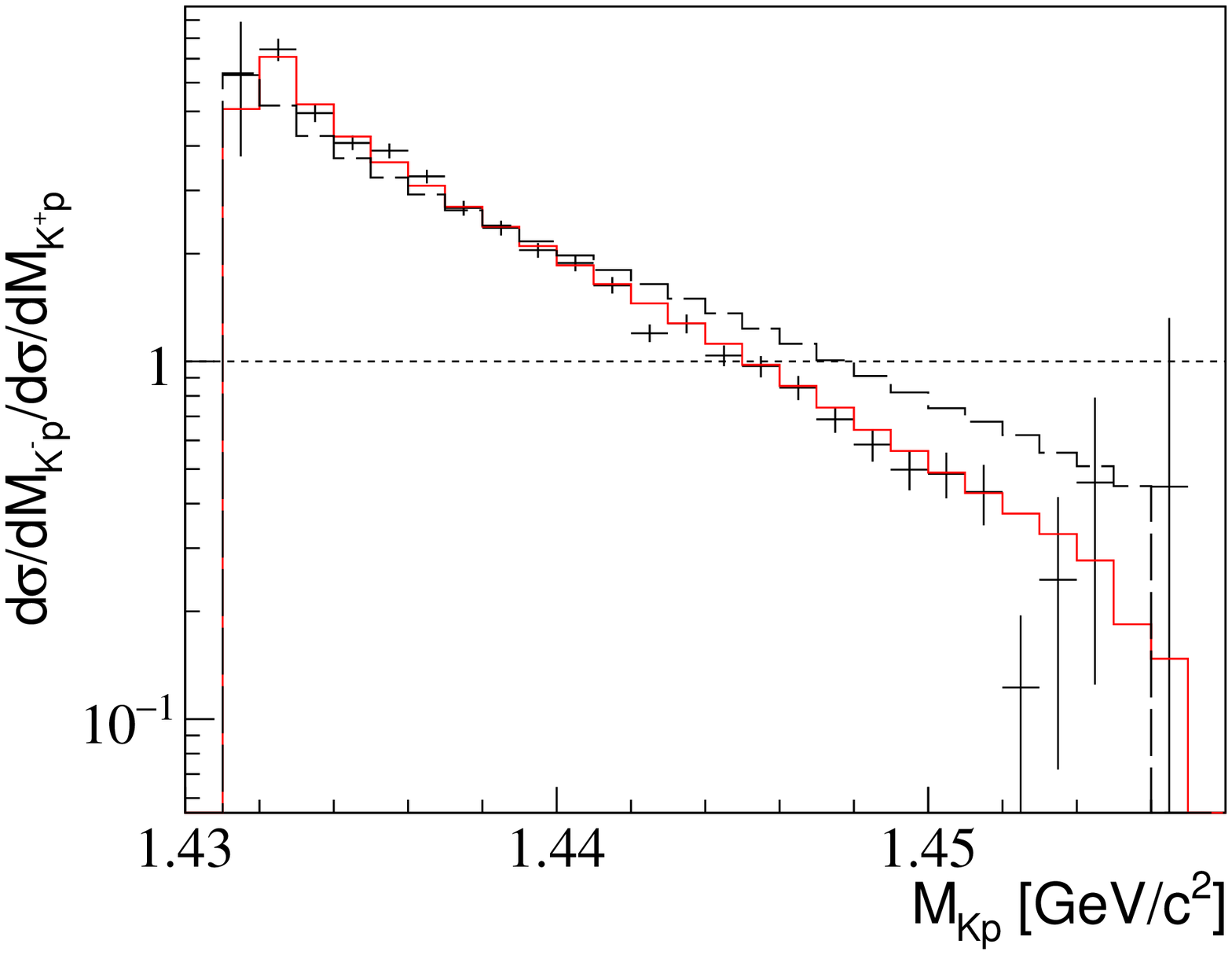}
\includegraphics[width=0.45\textwidth]{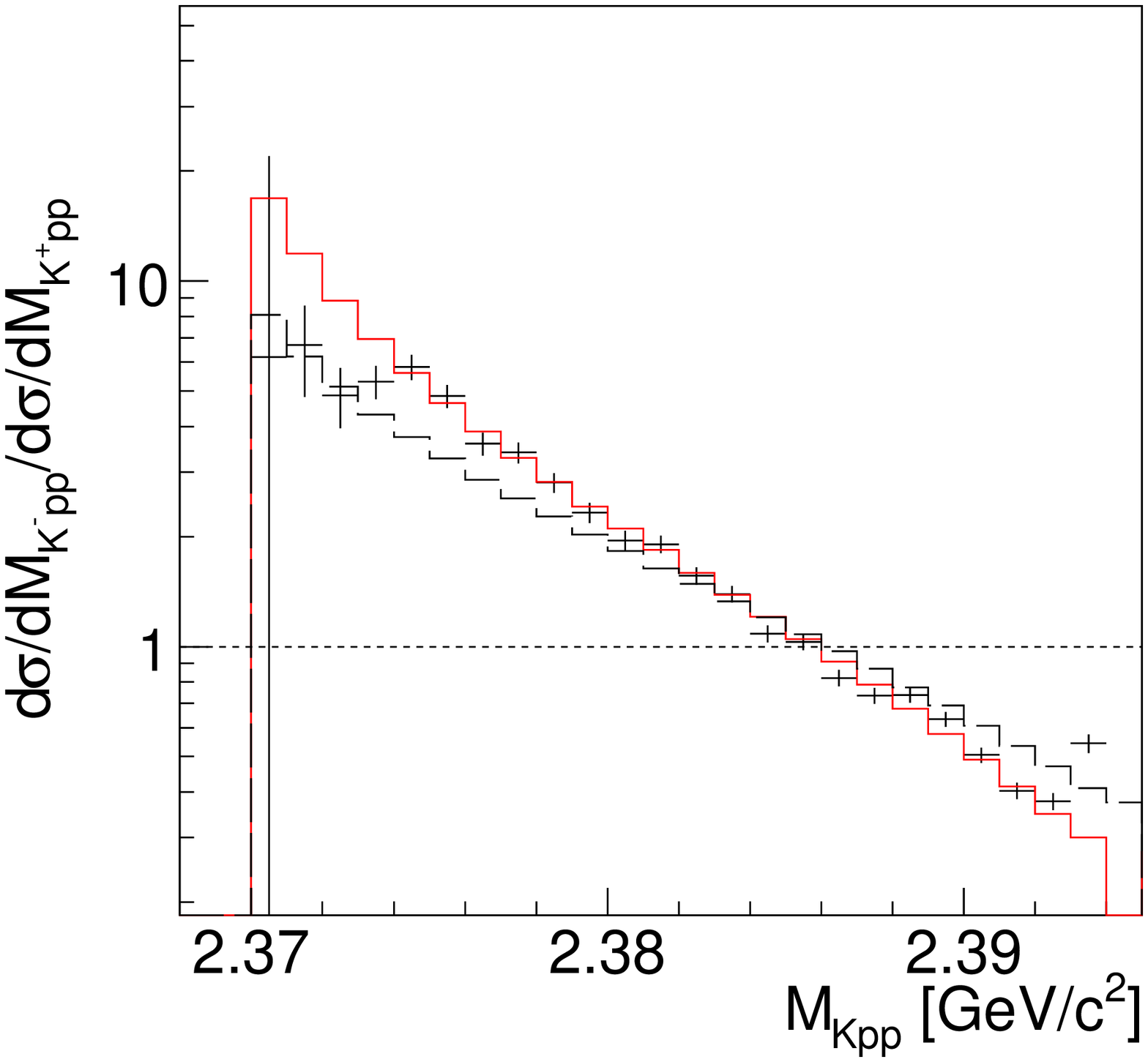}
\caption{\label{fig:Ye1} The measured ratios $R_{Kp}$ and $R_{Kpp}$ at
$Q=24$~MeV~\cite{Ye_13}. The red solid and broken black histograms represent
estimations that take into account $K^-p$, $pp$ and $K^+K^-$ final state
interactions with $a_{K^-p}=2.45i$~fm and $a_{K^-p}=1.5i$~fm, respectively. }
\end{center}
\end{figure}

Several experiments have shown that the $K^-$ is strongly attracted to one or
both protons in the $pp\to ppK^+K^-$ reaction. To put this on a quantitative
basis, define cross section ratios in terms of the $K^{\pm}p$ and $K^{\pm}pp$
invariant masses:
\begin{equation}
\label{IMratio}
R_{Kp} = \frac{d\sigma/dM_{K^-p}}{d\sigma/dM_{K^+p}}, \hspace{5mm}
R_{Kpp} = \frac{d\sigma/dM_{K^-pp}}{d\sigma/dM_{K^+pp}}\cdot
\end{equation}
The distributions in $R_{Kp}$ and $R_{Kpp}$ obtained in an experiment below
the $\phi$ threshold~\cite{Ye_13} are both shown in Fig~\ref{fig:Ye1}. The
data are well described with an effective scattering length of
$a_{K^-p}=2.45$~fm. Although $a_{K^-p}$ was taken to be purely imaginary, the
data are not very sensitive to the phase.

\subsection{Summary of information from semi-inclusive measurements}

Both the $\eta$ and the $K^-$ seem to be strongly attracted to protons at low
energy. This is not a complete surprise because there are $s$-wave resonances
sitting at (and overlapping with) the $\eta p$ and $K^-p$ thresholds. There
are thus strong $s$-wave couplings for $N^*(1535):\eta p$ and
$\Lambda(1405):K^-p$.

Since the $\eta$ is isoscalar, this means that the meson is also attracted to
neutrons. However, data on $pn \to dK^+K^-$~\cite{Maeda_09} can be
interpreted as suggesting that the $K^-$ attraction to neutrons is weaker
than to protons, probably because there is no $I=1$ $s$-wave hyperon
resonance near the $K^-n$ threshold. There is no firm evidence for strong
$\eta^{\prime}$, $\omega$, or $\phi$ attraction to nucleons at low energies
but the $\omega$ case is complicated by the decay width
$\Gamma_{\omega}\approx 8.5$~MeV.

\subsection{Information from inclusive experiments}

\begin{figure}[thb]
\begin{center}
\epsfig{file=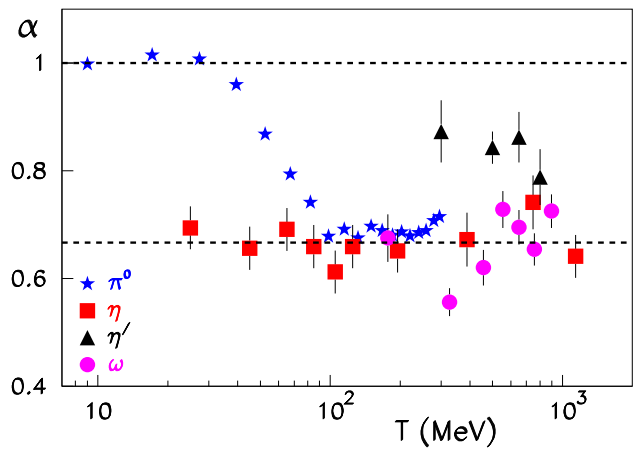,scale=0.5}
\caption{Scaling parameter $\alpha$ as a function of meson kinetic energy $T$
for $\pi^o$ \cite{Krusche_04}, $\eta$ \cite{Roebig_96,Mertens_08},
$\eta^{\prime}$ \cite{Nanova_12}, and $\omega$ mesons \cite{Kotulla_08}. Pion
data were not plotted above 300~MeV. \label{alpha}}
\end{center}
\end{figure}

Cross sections for inclusive photoproduction of a meson from a nucleus with
mass number $A$ are often fitted with $\sigma\sim A^{\alpha}$. If $\alpha$ is
close to unity, the whole nucleus is participating and the meson interaction
is weak. If $\alpha$ approaches 2/3, only the back surface is contributing
and the interaction is very strong. Data in Fig.~\ref{alpha} show that the
nucleus is fairly transparent to low energy pions, but the picture changes
when the $\Delta(1232)$ is reached. On the other hand, the $\eta$ is strongly
absorbed at all the energies shown.

The $\eta^{\prime}$ is an intermediate case but the $\omega$ also looks to be
strongly absorbed. However, only in the $\pi$ or possibly the $\eta$ case do
the data extend into the near-threshold s-wave domain. There are also
complications from two-step processes, e.g., $\gamma N \to \pi N$ followed by
$\pi N \to \omega N$. These might actually be kinematically favoured because
they share the momentum transfer better.

The $A^{\alpha}$ parameterisation is very close to that of the nuclear
transparency approach, where one compares the production on a nucleus to that
on a reference nucleus, which is invariably carbon, to form the ratio
\begin{equation}
R=\left(\frac{12}{A}\right)\,\frac{\sigma_A}{\sigma_{\rm C}}\cdot
\end{equation}

There are measurements of $R$ also in proton-nucleus collisions, e.g., in the
production of the $\phi$ meson~\cite{Polyankiy_11}. However, these are all
for $T_{\phi} > 160$~MeV and have little relevance for the mesic nucleus
question.

All the models used to analyse such data have large contributions from
two-step processes involving, perhaps, intermediate $\pi$ or $\omega$ mesons.
Hence the interpretation depends on the models used and it is difficult to be
sure from these how absorptive the $s$-wave $\phi p$ interaction really is.
Nevertheless, this is one of the few ways of getting some information
regarding the imaginary part of the potential between the meson and the
nucleus.

It has been suggested that the energy dependence of, say, $\gamma A \to
\eta^{\prime} A^*$ in and below the free nucleon threshold should be
sensitive to the $\eta^{\prime} A^*$ potential~\cite{Nanova_13}. There will
always be production below this threshold due, e.g., to Fermi motion. But, if
the $\eta^{\prime}$ is attracted to the nucleons, that effectively reduces
the total mass of a cluster and so this might be produced at lower photon
energies. A lot of corrections are included in the modeling but it is not
evident that all the uncertainties arising from the assumptions in the model
are taken into account. The distinction between the production on nucleon
clusters and two-step processes is not always clear and two-step effects
depend critically upon the particular meson produced.

\section{Production rates}

In order to form a mesic nucleus, a good production rate is needed as well as
a strong attraction of the meson to a nucleus. Though some information on
relative production rates is contained in the $pp \to ppX$ data, the $pd \to
{}^{3}{\rm He}X$ rates might be more informative. Define the average
amplitude squared by
\begin{equation}
|f(pd \to {}^{3}{\rm He}X)|^2 =
\frac{p_d}{p_X}\,\frac{\dd\sigma}{\dd\Omega}(pd \to {}^{3}{\rm He}X).
\end{equation}
Close to their thresholds
$|f(pd \to {}^{3}{\rm He}\eta)|^2 \approx 2500$~nb/sr, %
$|f(pd \to {}^{3}{\rm He}\omega)|^2 \approx 30$~nb/sr, %
$|f(pd \to {}^{3}{\rm He}\eta^{\prime})|^2 \approx 0.9$~nb/sr, %
$|f(pd \to {}^{3}{\rm He}\phi)|^2 \approx 2.3$~nb/sr.%

Thus $\eta^{\prime}$ production is more than three orders of magnitude weaker
than for the $\eta$, compared to a mere factor of twenty in $pp$ collisions.
The smallness of $\eta^{\prime}$ production seems to be confirmed by
unpublished COSY-WASA data~\cite{Wolke_12}. If the acceptance of WASA is
estimated on the basis of phase space then
$|f(pd \to {}^{3}{\rm He}\omega)|^2 \approx 11$~nb/sr at $Q=240$~MeV and %
$|f(pd \to {}^{3}{\rm He}\eta^{\prime})|^2 \approx 0.6$~nb/sr at $Q=64$~MeV.
Of course these are rough estimates but one would not be saved by a
factor of two in the $\eta^{\prime}$ case.

\section{Estimation of binding energies}

Liu and Haider~\cite{Liu_86} started the whole bound $\eta$-mesic business
through their estimates of binding within single-channel potential models,
where $V_{\eta A} \propto f_{\eta N}\,\rho(r)$, with $\rho(r)$ being the
nuclear density and $f_{\eta N}$ the $\eta$-nucleon elastic scattering
amplitude. This leasves several problems:

\begin{itemize}%
\item One does not know what to assume for $f_{\eta N}$. %
\item Due to the $N^*(1535)$ resonance, the potential is likely to have a
    strong energy dependence. How can this be taken into account? Which
    energy should one choose? %
\item It seems as though there may be nearby poles in the $\eta{}^3$He
    and $\eta{}^4$He systems. Who would trust the predictions of a
    one-particle optical potential for such light nuclei? %
\item Use of such potentials suggest that the binding energy of a meson
    to the ground state of a nucleus is likely to be similar in magnitude
    to that of one of its excited states, so that:
\[M(^{12}_{\eta}\textrm{C}(2^+)) - M(^{12}_{\eta}\textrm{C}(0^+)) \approx
4.4~\textrm{MeV}.\] This means that one would need very favourable
      kinematic conditions if the mesic nuclear widths were as large as
      the nuclear level spacing. It should be noted that the situation is
      not saved by neglecting states above the break-up threshold because
      the $\eta$ could just as well stick to one of the nuclear residues.
\end{itemize}

A Japanese group~\cite{Jido_12} has made estimates of the binding of an
$\eta^{\prime}$ to $^{12}$C in an optical potential approach for a variety of
potential strengths. One could argue that their potential is too attractive,
given what we know about the $\eta^{\prime}$-nucleon interaction, but the
widths that they predict are large compared to the nuclear level spacing. The
nuclear excited states problem is therefore likely to hinder most
$\eta^{\prime}$ mesic nuclei searches.

\section{Mesic nucleus experiments}

There are two very different methods to search for mesic nuclei:
\begin{enumerate}\item
Measure meson production at a few energies just above threshold and
attempt to extrapolate to below threshold, where a quasi-bound nucleus
may reside. This approach does overcome the very serious background
problem but it could only work if the mesic nucleus were lightly bound.
Even more troublesome is the fact that above-threshold experiments can
never distinguish between bound and virtual (antibound) states. This is
just like asking if one can deduce that the $^{3\!}S_1$ $np$ has a bound
state (deuteron) but the $^{1\!}S_0$ has none if one only looks above the
np threshold. A typical (i.e., best) example is $dp
\to{}^3\textrm{He}\eta$.%
\item Look directly in the bound state region. By definition the meson
    cannot emerge and the background could be overwhelming unless one
    could identify the quasi-free decay of the meson. But this is the
    only way to be 100\% sure that one has a quasi-bound state. One tries
    to suppress the background by choosing ``favourable'' kinematics. A
    typical example of this in the $\eta$ case is
    $_{\eta}^3\textrm{He}\to \pi^0 p X$, where the $\pi^0$ and $p$ come
    out back-to-back in the overall c.m.\ frame.
\end{enumerate}

\subsection{An above-threshold search}

There are a lot of data on $dp \to{}^3\textrm{He}\,\eta$. The total cross
section jumps to its plateau value within about 0.5~MeV of threshold. The
jump is even sharper if the beam momentum distribution is taken into
account~\cite{Mersmann_07}. There is a pole in the $\eta{}^3$He scattering
amplitude at $p_{\eta} = (-5 \pm 7) \pm i(19\pm 3)$~MeV/$c$, i.e., at $Q =
(-0.30\pm0.15) \pm i(0.21\pm0.06)$~MeV. Of course the real part can even
vanish by chance, but why is the imaginary part so small?

If the pole is due to the $\eta{}^3$He FSI, it should be present for all
entrance channels. A big near-threshold jump is seen in
$\gamma{}^3\textrm{He} \to \eta{}^3\textrm{He}$ but the resolution is not as
good as in the hadronic experiments~\cite{Pheron_12}. The best proof of the
FSI hypothesis is the deuteron tensor analysing power $T_{20}$ in $\pol{d}p
\to{}^3\textrm{He}\,\eta$, which is sensitive to the spin-3/2/spin-1/2 ratio
in the initial $dp$ state. The value of $T_{20}$ is effectively constant near
threshold~\cite{Papenbrock_14}, despite the cross section jumping around.

\subsection{A sub-threshold search}

Most sub-threshold (i.e., direct) searches have given disappointing results.
In the case of $\gamma{}^3\textrm{He} \to \pi^0 p X$, if this passes through
an $_{\eta}^{3}$He mesic nucleus, there could be a peak in the energy
distribution of back-to-back $\pi^0 p$ pairs in the overall c.m.\ frame. The
first MAMI experiment~\cite{Pfeiffer_04} found evidence for such a peak by
subtracting data in one angular bin from another. Later MAMI
data~\cite{Pheron_12} confirmed the existence of peak but showed that the
interpretation was very ``suspect''. The energy dependence showed lots of
structure but this seemed to evolve smoothly with $\pi^0 p$ opening angle.
There was no sign of any mesic nucleus decay.

\subsection{Quasi-bound $K^-pp$ systems}

The $K^+/K^-$ distortions in the COSY $pp \to K^+K^-pp$
data~\cite{Hartmann_06,Ye_13} seem to be driven mainly by $pp \to
K^+p\Lambda(1405)$ and the decay of the tail of $\Lambda(1405)\to
K^-p$~\cite{Xie_10}. This  might be an indication for a lightly bound
$K^-pp$, or $\Lambda(1405)p$, system, which could correspond to an $S = -1$,
$B=2$ mesic nucleus. Such a mesic nucleus could decay via $\Sigma^0\pi^0p$,
but counting rates for $pp \to K^+\Sigma^0\pi^0p$ are not very high and the
acceptance of the available spectrometers for multiparticle final states are
rather low!

There is a severe lack of experimental data for other nuclei. COSY studied
$pd \to{} ^3\textrm{He} K^+K^-$, but there was no magnetic field for the kaon
detection so that they could only produce average $K^{\pm}{}^3$He
distributions, which are little sensitive to the $K^-{}^3$He interaction.

But are there systems that are bound so deeply that they can only decay via
hyperon production? The subject is \underline{VERY} controversial!

\subsection{Deeply bound $K^-pp$ systems}

Yamazaki et al.~\cite{Yamazaki_10} took the 2.85~GeV DISTO 
data~\cite{Maggiora_01} and divided by a phase-space distribution that was
passed through the DISTO analysis program. They generated a $\Lambda p$
invariant-mass peak with $M \approx 2267$~MeV/$c^2$ and $\Gamma \approx
118$~MeV/$c^2$. This could be interpreted as a $\Lambda(1405)p$ bound state,
which they called $X(2267)$. However, there was no sign of such a state in
the 2.5~GeV DISTO data.

The procedure was studied in detail by Epple and Fabbietti~\cite{Epple_15},
who analysed the 3.5~GeV HADES data~\cite{Agakishiev_15}. They found that the
generated shape depended on the cuts imposed because the data just did not
look like phase space. They also questioned Yamazaki's estimate of the energy
dependence of $pp \to p\Lambda(1405)$, which was supposed to be the doorway
to his X(2267) state. Their criticism is described in detail in the Fabbietti
contribution to this workshop, to which the reader is
referred~\cite{Epple_15}.

%
%

\end{document}